\documentclass[groupedaddress,showpacs,showkeys,amssymb,eqsecnum,aps]{revtex4}     
 
\usepackage[pdftex]{graphicx}
\usepackage{amsmath} 
\usepackage{epsf}                                                                                           
\usepackage{color}                   
\usepackage{verbatim}                                                                         

\newcommand{\be}{\begin{equation}}                                                                          

\newcommand{\undertilde}[1]{\underset{\widetilde{}}{#1}}
\newcommand{\unf}{\undertilde{\; F}}
\newcommand{\unfa}{\unf\!{}_1}
\newcommand{\unfat}{\unf\!{}_1^{\;\rm T}}
\newcommand{\unfb}{\unf\!{}_2}
\newcommand{\unfbt}{\unf\!{}_2^{\;\rm T}}
\newcommand{\unfi}{\unf\!{}_i}

\newcommand{\una}{\undertilde{A}}
\newcommand{\unat}{\undertilde{A}^{\rm T}}

\newcommand{\und}{\undertilde{D}}

\newcommand{\unn}{\undertilde{\; N}}

\newcommand{\unv}{\undertilde{V}}
\newcommand{\unx}{\undertilde{X}}

\newcommand{\vphi}{\boldsymbol\phi}
\newcommand{\vphit}{\vphi^{\rm T}}
\newcommand{\vthm}{\boldsymbol{\theta}_{-}}
\newcommand{\vthmt}{\vthm^{\rm T}}
\newcommand{\vth}{\boldsymbol{\theta}}
\newcommand{\vp}{\boldsymbol{p}}
\newcommand{\vj}{\boldsymbol{j}}

\newcommand{\ee}{\end{equation}}

\newcommand{\dd}{{\rm d}^d x}

\newcommand{\imineq}[2]{\vcenter{\hbox{\includegraphics[height=#2ex]{#1}}}}
\begin{document}  

\title{Quantizing the Palatini Action using a Transverse Traceless Propagator}
\author{F. T. Brandt}
\email{fbrandt@usp.br}
\affiliation{Instituto de F\'{\i}sica, Universidade de S\~ao Paulo, S\~ao Paulo, SP 05508-090, Brazil}
\author{D. G. C. McKeon}
\email{dgmckeo2@uwo.ca}
\affiliation{
Department of Applied Mathematics, The University of Western Ontario, London, ON N6A 5B7, Canada}
\affiliation{Department of Mathematics and Computer Science, Algoma University,
Sault St.Marie, ON P6A 2G4, Canada}
\author{Chenguang Zhao}
\email{cheungylchaos@gmail.com}
\affiliation{
Department of Applied Mathematics, The University of Western Ontario, London, ON N6A 5B7, Canada}
                                                                                                              
\date{\today}
                                                                                                              
\begin{abstract}                                                                                              
We consider the first order form of the Einstein-Hilbert action and quantize it using the path integral.
Two gauge fixing conditions are imposed so that the graviton propagator is both traceless and transverse.
It is shown that these two gauge conditions result in two complex Fermionic vector ghost fields 
and one real Bosonic vector ghost field. All Feynman diagrams
to any order in perturbation theory can be constructed from two real Bosonic fields, two Fermionic ghost fields
and one real Bosonic ghost field that propagate. These five fields interact through just five three point vertices 
and one four point vertex.
\end{abstract}                                                                                                
                                                                       
\pacs{11.15.-q}
%PACS No.: 11.15.-q \\                                                                                                        
%KEY WORDS: gauge theories, first order, perturbation theory                                             
\keywords{gauge theories; first order; perturbation theory}
                                                               
\maketitle                     

%\coffeeA

%\newpage

\section{Introduction}
It has been shown with both Yang-Mills (YM) action and the Einstein-Hilbert (EH) action for gravity, 
that by using the first order form of the action, there is only a single vertex arising from the 
classical action and this is independent of momentum \cite{Okubo:1979gt,Buchbinder:1983ys,McKeon:1994ds,Kalmykov:1994fm,Brandt:2015nxa}.
This simplifies the computation of loop diagrams, even though the number of propagating fields is
increased.

It has also been shown that imposing both the conditions of tracelessness and transversality on the
spin two propagator associated with the EH action requires use of a non-quadratic gauge fixing 
Lagrangian \cite{Brandt:2007td,Brandt:2009qi,Brandt:2009rq,Brandt:2011zb,McKeon:2014iea}. 
Such gauge fixing results in the need to consider the contributions of two complex
Fermionic ghosts and one real Bosonic ghost analogous to the usual complex ``Faddeev-Popov'' ghosts.

In this paper we consider how the full first order Einstein-Hilbert (1EH) action can be used in conjunction with
the transverse-traceless (TT) gauge. We will show that the spin two propagator is TT only if the 
gauge fixing parameter $\alpha$ is allowed to vanish. This limit for $\alpha$ results in a well defined set
of Feynman rules with two propagating Bosonic fields, two complex Fermionic ghost fields, one real Bosonic ghost,
three three-point vertices for the Bosonic fields and four ghost vertices.

%The Ward identities (WTST identities) for the 1EH action in the TT gauge are considered in the
%appendix, as is independence of results on the gauge choice.

\section{The TT gauge for the 1EH Action}
The Einstein-Hilbert action in first order (Palatini) form
\begin{equation}\label{eq1}
S = \int \dd  \sqrt{-g} g^{\mu\nu} R_{\mu\nu}(\Gamma)
\end{equation}
when written in terms of the variables
\begin{subequations}\label{eq2}
\begin{equation}\label{eq2a}
h^{\mu\nu} = \sqrt{-g} g^{\mu\nu} 
\end{equation}
\mbox{and}
\be \label{eq2b}
G^\lambda_{\mu\nu} = \Gamma^\lambda_{\mu\nu} - \frac 1 2 \left(
\delta^\lambda_\mu \Gamma^\sigma_{\nu\sigma} + 
\delta^\lambda_\nu \Gamma^\sigma_{\mu\sigma} \right)
\ee
\end{subequations}
becomes
\begin{equation}\label{eq3}
S  = \int \dd   h^{\mu\nu}\left(G^\lambda_{\mu\nu\, ,\lambda}   +
\frac{1}{d-1} G^\lambda_{\lambda\mu}  G^\sigma_{\sigma\nu} -
G^\lambda_{\mu\sigma}  G^\sigma_{\nu\lambda} \right) .
\end{equation}
This ``Palatini'' form of the action facilitates a canonical analysis of $S$ \cite{McKeon:2010nf}.
It is equivalent to the second order form of the EH action at both the classical and quantum
levels \cite{Brandt:2015nxa}.
The diffeomorphism invariance of $S$ in Eq. \eqref{eq1} leads to the local gauge transformations
\begin{subequations}\label{eq4}
\begin{equation}\label{eq4a}
\delta h^{\mu\nu} = h^{\mu\lambda} \partial_\lambda \theta^\nu + h^{\nu\lambda} \partial_\lambda \theta^\mu -\partial_\lambda(h^{\mu\nu}\theta^\lambda)
\end{equation}
\begin{eqnarray}\label{eq4b}
\delta G^\lambda_{\mu\nu} &=& -\partial^2_{\mu\nu} \theta^\lambda
+\frac 1 2\left(\delta^\lambda_\mu\partial_\nu+\delta^\lambda_\nu\partial_\mu\right)\partial_\rho\theta^\rho-\theta^\rho\partial_\rho G^\lambda_{\mu\nu}
\nonumber \\ 
&+& G^\rho_{\mu\nu} \partial_\rho\theta^\lambda - \left(G^ \lambda_{\mu\rho} \partial_\nu+G^ \lambda_{\nu\rho} \partial_\mu\right)\theta^ \rho
\end{eqnarray}
\end{subequations}
The term bilinear in $h$ and $G$ in Eq. \eqref{eq3} does not lead to a well defined propagator, 
irrespective of the choice of gauge fixing. However, upon making an expansion of $h^{\mu\nu}$ 
about a flat background
\be\label{eq5}
h^{\mu\nu} = \eta^{\mu\nu} + \phi^{\mu\nu}(x)\;\;\; (\eta^{\mu\nu} = \mbox{diag}(+---\dots))
\ee
the term bilinear in $\phi$ and $G$ arising from Eq. \eqref{eq3} does have a well defined propagator
once an appropriate gauge fixing is chosen. These bilinear terms are the first order form of the
action for a spin two field \cite{McKeon:2010nf}.

In order to have a TT propagator for the spin two field we must consider a general gauge fixing
Lagrangian that is not quadratic \cite{Brandt:2007td}. If the classical Lagrange density appearing
in Eq. \eqref{eq3} is ${\cal L}(h^{\mu\nu},G^\lambda_{\mu\nu})$, then this entails inserting into the
generating functional
\be\label{eq6}
Z[j_{\mu\nu},J_\lambda^{\mu\nu}] = 
\int {\cal D} \phi^{\mu\nu} {\cal D} G^\lambda_{\mu\nu} 
\exp i\int \dd  \left({\cal L}(\eta+\phi,G) + j_{\mu\nu} \phi^{\mu\nu} + J^{\mu\nu}_\lambda G^\lambda_{\mu\nu}
\right) 
\ee                   
two factors of ``$1$'' 
\be\label{eq7}
1 = \int {\cal D} \vth_i \, \delta\left(\unf\!{}_i(\vphi + \una \vth_i) - \vp_i\right)
\det(\unf\!{}_i \una);\;\;\;  (i=1,2)
\ee
where $\vphi=(\phi^{\mu\nu}, G^\lambda_{\mu\nu})$. The gauge transformations of Eq. \eqref{eq4} 
are of the form
\be\label{eq8}
\delta_i \vphi = \una \vth_i
\ee
and the gauge fixing conditions are
\be\label{eq9}
\unf\!{}_i \vphi = 0.
\ee
Insertion of a third factor of ``$1$'' that is of the form
\be\label{eq10}
1 = \frac{1}{(\pi\alpha)^d}\int {\cal D} \vp_1 {\cal D} \vp_2 \exp
\frac{-i}{\alpha}\int \dd  (\vp_1^{\rm T} \unn \vp_2) \det(\unn) 
\ee
into Eq. \eqref{eq6} leads to
\begin{eqnarray}\label{eq11}
Z[\vj] = \int {\cal D}\vphi \det(\unf\!{}_1  \una) \det(\unf\!{}_2 \una)
\det(\unn/\pi\alpha) \int {\cal D} \vth_1 {\cal D} \vth_2 \nonumber \\
\exp i\int \dd \left\{
{\cal L}(\vphi) - \frac{1}{\alpha} \left[\unf\!{}_1(\vphi+\una\vth_1)\right]^{\rm T}%\!\!
\unn \left[\unf\!{}_2(\vphi+\una\vth_2)\right] + \vj^{\rm T} \cdot \vphi\right\} ;\;\;\; 
(\vj\equiv (j_{\mu\nu}, J_\lambda^{\mu\nu})) .
\end{eqnarray}

Since the gauge transformation of Eq. \eqref{eq8} leaves ${\cal L}(\vphi)$, 
${\cal D}\vphi$ and $\det(\unf\!{}_i\una)$ invariant \cite{faddeev:book1980,weinberg:book1995},
we can make the shift
\begin{equation}\label{eq12a}
\vphi \rightarrow\vphi -  \una (\vth_{+} + \epsilon \vth_{-})
\end{equation}
in Eq. \eqref{eq11} \hbox{($\vth_{\pm} \equiv (\vth_1\pm\vth_2)/2$)} 
leaving us with 
\begin{eqnarray} \label{eq12}
Z[\vj] = \int {\cal D}\vphi {\cal D}\vth_{-} 
\det(\unf\!{}_1 \una) \det(\unf\!{}_2 \una) \det(\unn) %/\pi\alpha) 
\nonumber \\
\exp i\int \dd \left\{
{\cal L}(\vphi) - \frac{1}{\alpha} \left[\unf\!{}_1(\vphi+\una(1-\epsilon)\vth_{-})\right]^{\rm T}
\right.\nonumber \\
\left. 
\unn \left[\unf\!{}_2(\vphi-\una(1+\epsilon)\vth_{-})\right] + \vj^{\rm T} \cdot \vphi\right\}. 
\end{eqnarray}
A factor $1/(\pi \alpha)^{d/2} \int {\cal D}\vth_{+}$ has been absorbed into the normalization
of $Z$. We now choose the gauge fixing to be
\begin{subequations}\label{eq13}
\be
\unf\!{}_i \vphi = g_i \partial_\rho \phi^\mu_\mu + \partial_\mu \phi^\mu_\rho
\ee
and
\be
\unn = {\eta^{\mu\nu}}/{2}.
\ee
\end{subequations}
%%%%%%%%%%%%%%%%%%%%%%%%%%%%%%%%%%%%  Insert 1   %%%%%%%%%%%%%%%%%%%%%%%%%%%%%%%%%%%%%
The gauge fixing contribution of Eq. \eqref{eq12} becomes
\begin{eqnarray}\label{eq14}
&&\left[\unf\!{}_1(\vphi+\una(1-\epsilon)\vthm)\right]^{\rm T}
\unn \left[\unf\!{}_2(\vphi-\una(1+\epsilon)\vthm)\right] 
\nonumber\\
&=& (\unf\!{}_1\vphi)^{\rm T} \unn (\unf\!{}_2\vphi)+(\epsilon^2-1)\left\{\left[\vthmt +
\frac 1 2 \vphit\left(-(1+\epsilon) \unfat \unn \unfb +(1-\epsilon) \unfbt \unn \unfa\right) \una
\right.\right. \nonumber \\ 
&&\left.\left({(\unat \unfat\unn\unfb\una)^{-1}}/{(\epsilon^2-1)}\right)\right]
\left[\unat\unfat\unn\unfb\una\right]
\nonumber \\
&& \left. \left[\vthm +\frac 1 2 \left((\unat\unfat\unn\unfb\una)^{-1}/(\epsilon^2-1)\right)
\unat\left(-(1+\epsilon)\unfbt\unn\unfa +(1-\epsilon)\unfat\unn\unfb\right)\vphi \right] \right\}
\nonumber \\
&& 
-\frac{1}{4(\epsilon^2-1)}\vphit\left(-(1+\epsilon)\unfat\unn\unfb + (1-\epsilon)\unfbt\unn\unfa\right)\una 
(\unat\unfat\unn\unfb\una)^{-1} \unat
\nonumber \\ &&
\left(-(1+\epsilon)\unfbt\unn\unfa + (1-\epsilon)\unfat\unn\unfb \right)\vphi  
\end{eqnarray}
(In Eq. \eqref{eq14} we use the convention $\partial^{\rm T} = -\partial$.)

Provided $\epsilon\neq\pm 1$, the shift in $\vthm$
\begin{equation}\label{eq15}
\vthm \rightarrow \vthm - \frac 1 2\left(
(\unat\unfat\unn\unfb\una)^{-1}/(\epsilon^2-1) \right)
\unat\left(-(1+\epsilon)\unfbt\unn\unfa +(1-\epsilon)\unfat\unn\unfb\right)\vphi
\end{equation}
can be made to diagonalize Eq. \eqref{eq14} in $\vthm$ and $\vphi$. In Refs. \cite{Brandt:2007td,Brandt:2009qi,Brandt:2009rq}
$\epsilon=\pm 1$ and a shift in $\vphi$ was used to diagonalize the gauge fixing, but as such a shift is not a gauge transformation,
${\cal L}(\vphi)$ is not invariant under this transformation and new vertices involving $\vphi$ and $\vthm$ must be
introduced. We take $\epsilon\neq\pm 1$ in order to be able to make a shift in $\vthm$ that eliminates mixed propagators for these fields
without introducing extra vertices.

Together Eqs. \eqref{eq14} and \eqref{eq15} result in
\begin{eqnarray} \label{eq17}
Z[\vj] &=& \int {\cal D}\vphi {\cal D}\vth_{-} 
\det(\unf\!{}_1 \una) \det(\unf\!{}_2 \una) \det(\unn) %/\pi\alpha) 
\nonumber \\
&&\exp i\int \dd \left\{
{\cal L}(\vphi) - \frac{1}{\alpha} (\unfa\vphi)^{\rm T} \unn(\unfb\vphi) 
\right. 
%\nonumber \\
%&&
-\frac{1}{\alpha(\epsilon^2-1)}\vthmt(\unat\unfat\unn\unfb\una)\vthm 
\nonumber \\
&+&\frac{1}{4\alpha(\epsilon^2-1)}\vphit\left(-(1+\epsilon)\unfat\unn\unfb + (1-\epsilon)\unfbt\unn\unfa\right)\una 
(\unat\unfat\unn\unfb\una)^{-1}
\nonumber \\
&&\left.  \unat\left(-(1+\epsilon)\unfbt\unn\unfa + (1-\epsilon)\unfat\unn\unfb \right)\vphi  
+ \vj^{\rm T} \cdot \vphi\right\}. 
\end{eqnarray}
The integral over $\vthm$ can now be evaluated in Eq. \eqref{eq17}; it results in a contribution
\begin{equation}\label{eq18}
\det{}^{-1/2}(\unfa\una)\det{}^{-1/2}(\unn)\det{}^{-1/2}(\unfb\una).
\end{equation}
We now treat the last term in Eq. \eqref{eq17} as an interaction term. Due to its structure, the two fields $\vphi$ that occur
explicitly ($\una$ also is $\vphi$ dependent on account of Eq. \eqref{eq4a}) are contracted with a propagator for $\phi_{\mu\nu}$ 
and a factor of $\unx$ where
\begin{eqnarray}\label{eq19}
X_{\mu\nu,\lambda\sigma} &\equiv& \left(-(1+\epsilon)\unfbt\unn\unfa + (1-\epsilon)\unfat\unn\unfb\right)_{\mu\nu,\lambda\sigma} 
\nonumber\\
&=&\frac 1 2 (g_1-g_2)\left(\partial_\mu\partial_\nu\eta_{\lambda\sigma} - \eta_{\mu\nu}  \partial_\lambda\partial_\sigma \right)
\nonumber \\
&+& \epsilon\left[g_1 g_2 \eta_{\mu\nu} \eta_{\lambda\sigma} \partial^2 + \frac{g_1+g_2}{2}
\left(\partial_\mu\partial_\nu \eta_{\lambda\sigma} + \eta_{\mu\nu} \partial_\lambda\partial_\sigma \right) \right.
\nonumber \\
&+&\left. \frac 1 4 \left(\partial_\mu\partial_\lambda \eta_{\nu\sigma}  + \partial_\nu\partial_\lambda \eta_{\mu\sigma} +  
                         \partial_\mu\partial_\sigma \eta_{\nu\lambda}  + \partial_\nu\partial_\sigma \eta_{\mu\lambda}  \right)\right]
\end{eqnarray}
by Eq. \eqref{eq13}.

We know from Refs. \cite{Brandt:2007td,Brandt:2009qi,Brandt:2009rq} that as $\alpha\rightarrow 0$, the propagator for the field
$\phi_{\mu\nu}$ that comes from ${\cal L}(\vphi) -\frac{1}{\alpha} (\unfa\vphi)^{\rm T} \unn (\unfb\vphi)$ is transverse and traceless in the
limit $\alpha\rightarrow 0$ provided $g_1\neq g_2$. Only terms of order $\alpha$ are not transverse and traceless.
Thus, on account of the structure of Eq. \eqref{eq19}, the contribution of the vertex coming from the last term in Eq. \eqref{eq17} vanishes
as $\alpha\rightarrow 0$, even though this vertex is proportional to $1/\alpha$. There is on exception to this; when a sequence of these
vertices lies in a ring, then a finite contribution arises in the limit $\alpha\rightarrow 0$. To see this in more detail, write this last term in
Eq. \eqref{eq17} as
\begin{equation}\label{eq20}
\frac{1}{\alpha} \vphit \unv \vphi = \frac{1}{\alpha} \vphit (\unx^{\rm T}\una) \frac{(\unat\unfat\unn\unfb\una)^{-1}}{4(\epsilon^2-1)}\unat\unx\vphi .
\end{equation}
A ring in which a sequence of these vertices occurs results in a contribution proportional to
\begin{eqnarray}\label{eq21}
&{\rm Tr}\left\{\left[\frac{1}{\alpha} \unx^{\rm T} \una (\unat\unfat\unn\unfb\una)^{-1} \unat\unx \right]\und 
\left[\frac{1}{\alpha} \unx^{\rm T} \una (\unat\unfat\unn\unfb\una)^{-1} \unat\unx \right] \und  \right. \nonumber\\
&\dots \left .
\left[\frac{1}{\alpha} \unx^{\rm T} \una (\unat\unfat\unn\unfb\una)^{-1} \unat\unx \right] \und 
\right\},
\end{eqnarray}
where $\und$ is the propagator of $\vphi$. From Eq. \eqref{eq19} it is apparent that since when $\alpha\rightarrow 0$ $\und$ is transverse and traceless,
then $\unx\und$ is of order $\alpha$; since we have a factor of $1/\alpha$ for each factor of $\unx\und$ on account of these vertices occurring in a ring,
we can let
\begin{equation}\label{eq22}
\lim_{\alpha\rightarrow 0} \frac{1}{\alpha} \unx\und = \unx\und^{(0)}.
\end{equation}
Furthermore, a contribution of a closed loop of these vertices can be written as
\begin{eqnarray}\label{eq23}
&&\det{}^{-1/2}\left[\unx^{\rm T} \una (\unat\unfat\unn\unfb\una)^{-1} \unat\unx \und^{(0)} \right] 
\nonumber \\
&=&
\det{}^{1/2}(\unfa\una) \det{}^{1/2}(\unn) \det{}^{1/2}(\unfb\una) \det{}^{-1/2}(\unat\unx\und^{(0)}\unx^{\rm T}\una) .
\end{eqnarray}
Together Eqs. \eqref{eq18} and \eqref{eq23} reduce Eq. \eqref{eq17} to 
\begin{eqnarray}\label{eq24}
Z[\vj] &=& \lim_{\alpha\rightarrow 0}\int {\cal D}\vphi 
\det(\unf\!{}_1 \una)  \det(\unn)   \det(\unf\!{}_2 \una) \det{}^{-1/2}(\unat\unx\und^{(0)} \unx^{\rm T}\una) 
\nonumber \\
&&\exp i\int \dd \left\{
{\cal L}(\vphi) - \frac{1}{\alpha} (\unfa\vphi)^{\rm T} \unn(\unfb\vphi) + \vj^{\rm T} \cdot \vphi\right\}
\end{eqnarray}
provided $g_1\neq g_2$. The functional determinants in Eq. \eqref{eq24} can be exponentiated using ``ghost'' fields; 
$\det(\unfi\una)$ ($i=1,2$) using complex Fermionic ``Faddeev-Popov''
ghosts ${\bf c} _i$ \cite{Feynman:1963ax,DeWitt:1967yk, Faddeev:1967fc, Mandelstam:1968hz},
$\det(\unn)$ by a complex Fermionic Nielsen-Kalosh ghost \cite{Nielsen:1978mp, Kallosh:1978de} and
$\det{}^{-1/2}(\unat\unx\und^{(0)}\una)$ by a real Bosonic ghost ${\bf\zeta}$. By Eq. \eqref{eq4a}, it follows that
\begin{equation}\label{eq25}
(\una\vth)_{\mu\nu} = \left[\partial_\mu\eta_{\nu\rho} +\partial_\nu\eta_{\mu\rho} - \partial_\rho\eta_{\mu\nu} +
\left(\phi_\mu{}^\sigma\partial_\sigma \eta_{\nu\rho}+\phi_\nu{}^\sigma\partial_\sigma \eta_{\mu\rho}+\partial_\rho\phi_{\mu\nu}\right) \right]\theta^\rho .
\end{equation} 
Using Eqs. \eqref{eq19} and \eqref{eq25} and the propagator for $\vphi$ given in Ref \cite{Brandt:2007td} we find that the contribution that is bilinear in the ghost
${\bf\zeta}$is given by 
\begin{equation}\label{eq26}
4p^2\zeta_\mu\left\{\epsilon^2 p^2\eta^{\mu\nu} + \left[\left(g_1 g_2(d-2)^2 - (g_1+g_2)(d-2)\right)(\epsilon^2-1) -1\right]p^\mu p^\nu\right\}\zeta_\nu
\end{equation}
which becomes 
\begin{equation}\label{eq27}
4 p^4\epsilon^2\zeta_\mu\eta^{\mu\nu} \zeta_\nu .
\end{equation}
when
\begin{equation}\label{eq28}
g_1 = -g_2 = \frac{1}{(d-2)\sqrt{1-\epsilon^2}}.
\end{equation}
Similarly, the vertex for $\phi_{\mu\nu}(p)$ -- $\zeta_\alpha(q)$ -- $\zeta_\beta(r)$ comes from
\begin{eqnarray}\label{eq29}
&&\frac 1 2 \left\{\left[(d-2)g_1(\epsilon-1)^2 + (d-2) g_2 ((\epsilon+1)^2 -2\right]
q^\mu q^\alpha \left(p^\beta q^\nu + r^\beta q^\nu - 2 q^\beta r^\nu\right)   \right.
\nonumber \\
&+&  \epsilon^2 q^2 q^\mu \left[2 r^\nu\eta^{\alpha\beta} - p^\beta\eta^{\alpha\nu} + r^\beta\eta^{\alpha\nu}\right]
\nonumber \\
&+& q^2\left(2 r^\nu q^\alpha \eta^{\mu\beta} - p^\beta q^\alpha \eta^{\mu\nu} - r^\beta q^\alpha \eta^{\mu\nu}\right)
%\nonumber \\ && 
\left[g_1(\epsilon+1)^2+g_2(\epsilon-1)^2 - 2 g_1g_2(d-1)(\epsilon+1)^2\right]
\nonumber \\ &+& \left.
\epsilon^2 q^2 \eta^{\mu\nu}\left(2 r^\nu q^\beta - p^\beta q^\nu -r^\beta q^\nu \right) \right\}
+ (\mu\leftrightarrow\nu) + (\alpha\leftrightarrow\beta ;\; q\leftrightarrow r).
\end{eqnarray}
Finally, a vertex for $\phi_{\mu_1\nu_1}(p)$ -- $\phi_{\mu_2\nu_2}(q)$ -- $\zeta_\alpha(r)$ -- $\zeta_\beta(s)$ can also be worked out.
The vertices $\phi$ -- $\phi$ -- $\zeta$ -- $\zeta$  and \mbox{$\phi$ -- $\zeta$ -- $\zeta$} are both quartic in the external momenta.

%%%%%%%%%%%%%%%%%%%%%%%%%%%%%%%%% End of Insert 1 %%%%%%%%%%%%%%%%%%%%%%%%%%%%%%%%%%%%%%%%%%

The two complex ``Faddeev-Popov'' ghosts ${\bf c}_1$ and ${\bf c}_2$ and the real Bosonic ghost $\zeta$  reduce to a single
complex Fermionic Faddeev-Popov ghost ${\bf c} = {\bf c}_1 + i {\bf c}_2$ if we deal with a quadratic gauge fixing Lagrangian
when $\unfa = \unfb$.  

If we now define $M^{\mu\nu}_\lambda{}^{\pi\tau}_\sigma(h)$ by the equation 
\be \label{eq30}
h^{\mu\nu} \left(\frac{1}{d-1} G^\lambda_{\lambda\mu} G^\sigma_{\sigma\nu}
- G^\lambda_{\sigma\mu} G^\sigma_{\lambda\nu} \right)
=\frac 1 2 M^{\mu\nu}_\lambda {}^{\pi\tau}_\sigma(h) G^\lambda_{\mu\nu} G^\sigma_{\pi\tau}
\ee
then the shift
\be \label{eq31}
G^\lambda_{\mu\nu} \rightarrow G^\lambda_{\mu\nu}  + 
M^{-1}{}^\lambda_{\mu\nu}{}^\sigma_{\pi\tau}(\eta)\, \phi^{\pi\tau}_{,\sigma}
\ee
in ${\cal L}(\vphi)$ in Eq. \eqref{eq22} leads to 
\begin{eqnarray}\label{eq32}
{\cal L}(\vphi) = -\frac 1 2 
\phi^{\mu\nu}_{,\lambda} M^{-1}{}^\lambda_{\mu\nu}{}^\sigma_{\pi\tau}(\eta)\, \phi^{\pi\tau}_{,\sigma}
+ \frac 1 2 G^{\lambda}_{\mu\nu} M^{\mu\nu}_\lambda{}^{\pi\tau}_{\sigma}(\eta) G_{\pi\tau}^{\sigma}
\nonumber \\
+ \frac 1 2
\left(G^\lambda_{\mu\nu}  +  \phi^{\alpha\beta}_{,\xi}  M^{-1}{}^\xi_{\alpha\beta}{}^\lambda_{\mu\nu}(\eta)\right)
M^{\mu\nu}_\lambda{}^{\pi\tau}_{\sigma}(\phi) 
\left(G^\sigma_{\pi\tau}  +  M^{-1}{}^\sigma_{\pi\tau}{}^\zeta_{\gamma\delta}(\eta) \phi^{\gamma\delta}_{,\zeta}   \right)
\end{eqnarray}
so that off diagonal propagators $\phi-G$ are eliminated. However, two new momentum 
dependent vertices now arise. They are $\phi-\phi-\phi$ and $\phi-\phi-G$.

With the gauge fixing of Eq. \eqref{eq13} we find from Ref. \cite{Brandt:2007td} that the propagator for the field
$G^\lambda_{\mu\nu}$ is
\begin{subequations}\label{eq33}
\begin{eqnarray}%{lll}
\displaystyle{\substack{{}^{\displaystyle{ {}_{\mu\nu}^\lambda}} {\; \includegraphics[scale=0.7]{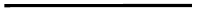} \;} {}^{\displaystyle{ {}_{\pi\tau}^\rho}} 
\\  {\displaystyle{ }}}} 
&  =  & \;\;\;\;
%{\cal D}_{G}{}^\lambda_{\mu\nu} {}^\rho_{\pi\tau} 
%
%{\cal D}_G{}_{\mu\nu}^{\lambda} \; {}_{\pi\tau}^{\rho}   &\equiv&   
%
\frac{1}{4} \eta^{\lambda \rho } \left(\eta_{\mu\tau } \eta_{\nu \pi} + \eta_{\mu \pi } \eta_{\nu \tau}
-\frac{2}{d-2} \eta_{\mu \nu } \eta_{\pi \tau }\right) 
\nonumber \\
&&
-\frac{1}{4} \left( \delta^{\lambda}_{ \tau } \delta_{\mu}^{ \rho } \eta_{\nu \pi}
+\delta^{\lambda}_{\tau } \delta_{\nu}^{ \rho}\eta_{\mu \pi } 
+\delta^{\lambda}_{  \pi }  \delta_{\nu}^{ \rho} \eta_{\mu \tau } 
+\delta^{\lambda}_{ \pi } \delta_{\mu}^{ \rho } \eta_{\nu \tau} \right) 
%\equiv {\cal D}{}_{\mu\nu}^\lambda{}_{\pi\tau}^\rho  
%
%{}^\lambda_{\alpha\beta} {}^\sigma_{\gamma\delta}
%\nonumber 
\end{eqnarray}
\mbox{The propagator for $\phi_{\mu\nu}$ is \cite{Brandt:2007td}}  
\begin{eqnarray}\label{eq33b}
\displaystyle{\substack{{}^{\displaystyle{ {}_{\displaystyle{\mu\, \nu}}}} 
{\; \includegraphics[scale=0.7]{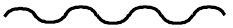} \;} {}^{\displaystyle{{}_{\displaystyle{\lambda\, \sigma}}}} 
\\  {\displaystyle{k}}}}
&  =  & \;\;\;\;
%{\cal D}_\phi^{\mu\nu\,\rho\sigma} =
\frac{1}{k^2}\left\{\eta_{\mu\lambda} \eta_{\nu\sigma} + \eta_{\mu\sigma} \eta_{\nu\lambda} 
- 2 \frac{(g_1-g_2)^2 + 2 (g_1+1)(g_2+1)\alpha}{\Delta} \eta_{\mu\nu} \eta_{\lambda\sigma}\right.
\nonumber\\   &+& 
(\alpha-1)\frac{1}{k^2}\left[k_\mu k_\lambda \eta_{\nu\sigma}
+(\mu\leftrightarrow\nu)+(\lambda\leftrightarrow\sigma)\right]
\nonumber\\   &+& 
2 \frac{(g_2-g_1)^2 + [4(g_1+1)(g_2+1)-g_2-g_1-2]\alpha}{\Delta} 
\frac{1}{k^2}\left[k_\mu k_\nu \eta_{\lambda\sigma} + k_\lambda k_\sigma \eta_{\mu\nu}\right]
\nonumber\\   &+& \frac{1}{\Delta} 
\left[4\alpha\left[(g_1+g_2)(d-4)+(2 g_1 g_2+1)(d-3)-\left(g_1^2+g_2^2\right)(d-1)\right]
\right. 
\nonumber \\ &+&\left. \left. 
2(d-2)\left[(g_1-g_2)^2-\alpha ^2 (4(g_1+1)(g_2+1)-1) \right]\right]
\frac{1}{k^4} k_\mu k_\nu k_\lambda k_\sigma \right\},
\end{eqnarray}
\mbox{where $\Delta=(d-1)(g_1-g_2)^2 + 2 (d-2)(g_1+1)(g_2+1)\alpha$. 
When $\alpha\rightarrow 0$ ($g_1\neq g_2$) this becomes the transverse-traceless}
\mbox{propagator.}

\mbox{For the real fields $c_i$ we have   }
\begin{eqnarray}%{lll}
\substack{ \displaystyle{\mu} \; \includegraphics[scale=0.7]{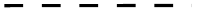} \; \displaystyle{\nu}
\\  {\displaystyle{ k}}}
& = &  D_{\mu\nu}^{(i)} \;\;\;\;
%{\cal D}^{gh}_{\mu\nu}(p)  
%{}_{\mu\nu}
= \frac{\displaystyle{\frac{(d-2)g_i k_\mu k_\nu}{k^2 [(d-2) g_i -1]}}  -\eta^{\mu\nu} }{k^2}.
%\nonumber
\end{eqnarray}
\end{subequations}
The vertices are given by
\begin{subequations}\label{eq34}
\begin{eqnarray}%{lll}
{}_{\mu\nu},\, p\, { \imineq{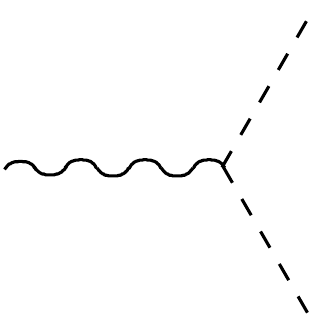}{14} }^{{\displaystyle{i,\; q,\;{}_\alpha}}}_{{{\displaystyle{j,\;r,\; {}_\beta\;\; }}}}
&  =  & \;\;\;\;
%{\cal V}^{gh}{\;\,}_{\mu\nu}^{\alpha\beta}(p,q,r) 
\frac{\delta^{ij} }{4}\left[-p^\beta q^\nu \eta^{\mu\alpha} -r^\beta q^\nu \eta^{\mu\alpha}
-g_i p^\beta q^\alpha \eta^{\mu\nu}  
\right. \nonumber \\
& - & \left. g_i r^\beta q^\alpha \eta^{\mu\nu}
+2  g_i r^\nu q^\alpha \eta^{\mu\beta} 
%\nonumber \\  & + & \left. 
 + (q,\alpha) \leftrightarrow (r,\beta) \right] + \mu\leftrightarrow\nu
%\nonumber
\end{eqnarray}
\begin{eqnarray}%{lll}
{}_{\mu\nu} { \imineq{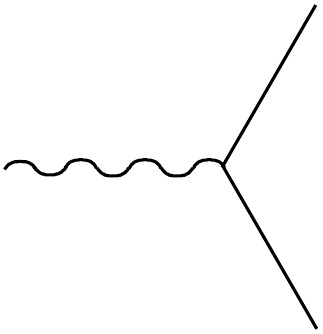}{14} }^{\displaystyle{\;\;{}^{\alpha\beta}_\lambda}}_{\displaystyle{\;\;{}^{\gamma\delta}_\sigma}}
&  =  & \;\;\;\;
%{\cal V} {}_{\mu\nu} {}_\lambda^{\alpha\beta}{}_\sigma^{\gamma\delta} 
%{}_{\mu\nu} {}^\lambda_{\alpha\beta}{}^\sigma_{\gamma\delta}
\frac{1}{8} \left\{\left[\left(
\frac{\delta_{\mu}^{\beta} \delta_{\nu}^{\delta} \delta^\alpha_\lambda \delta^\gamma_\sigma}{d-1} - 
\delta_{\mu}^{\beta} \delta_{\nu}^\delta \delta^\alpha_\sigma \delta^\gamma_\lambda + \mu \leftrightarrow \nu\right) + \alpha\leftrightarrow\beta 
\right]+\gamma\leftrightarrow\delta\right\}
\nonumber \\
&& + \;\; (\lambda, \alpha,\beta) \longleftrightarrow (\sigma,\gamma,\delta)
%\nonumber 
\end{eqnarray}
%\end{equation}
%\\
%\nonumber 
%\\
%\begin{equation}
\begin{eqnarray}%{lll}
{}_\sigma^{\gamma\delta}\, p\, { \imineq{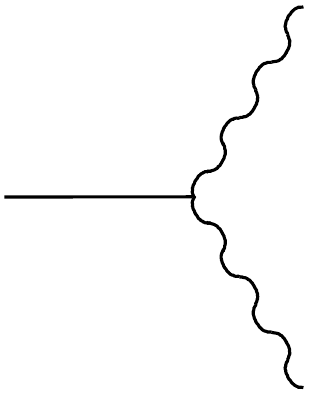}{14} }^{\displaystyle{q\;{}_{\mu\nu}}}_{\displaystyle{r\;{}_{\alpha\beta}}}
&  =  & \;\;\;\;
%{\cal V}^\lambda_{\mu\nu} {}_{\alpha\beta}{}_{\gamma\delta}(p,q,r) 
%{}_{\mu\nu} {}^\lambda_{\alpha\beta}{}^\sigma_{\gamma\delta}
\frac{i r_\theta}{4}\left\{\left[\left(
\frac{1}{d-1} \delta^\gamma_\mu \delta^\delta_\sigma {\cal D}_{ \alpha\beta}^\theta {}^\rho_{\nu\rho}
-\delta^\gamma_\mu {\cal D}_{\alpha\beta}^\theta{}_{\nu\sigma}^\delta  
+ \mu \leftrightarrow \nu\right) + \alpha\leftrightarrow \beta\right] + \gamma\leftrightarrow \delta\right\}
\nonumber \\
&& +  \;\; (q, \alpha,\beta) \longleftrightarrow (r,\mu,\nu)
%\nonumber 
\end{eqnarray}
%\end{equation}
%\\
%\nonumber 
%\\
%\begin{equation}
\begin{eqnarray}%{lll}
{}_{\mu\nu}\, p\, { \imineq{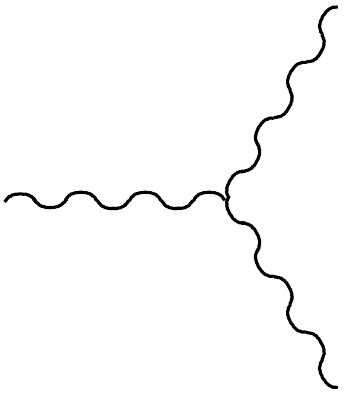}{14} }^{\displaystyle{q\;{}_{\alpha\beta}}}_{\displaystyle{r\;{}_{\gamma\delta}}}
&  =  & \;\;\;\;
%{\cal V} {}_{\mu\nu} {}_{\alpha\beta}{}_{\gamma\delta} (p,q,r) 
%{}_{\mu\nu} {}^\lambda_{\alpha\beta}{}^\sigma_{\gamma\delta}
\frac{q_\kappa r_\theta}{8}\left\{\left[\left(
{\cal D}{}^\kappa_{\alpha\beta}{}^\pi_{\mu\sigma} 
                                                                {\cal D}{}^\theta_{\gamma\delta}{}^\sigma_{\nu\pi} 
 -\frac{1}{d-1} {\cal D}{}^\kappa_{\alpha\beta}{}^\sigma_{\mu\sigma} 
                                                                {\cal D}{}^\theta_{\gamma\delta}{}^\pi_{\nu\pi} 
 + \mu \leftrightarrow \nu\right) + \alpha\leftrightarrow \beta\right] + \gamma\leftrightarrow \delta\right\}
\nonumber \\
&& + \;\; \mbox{six permutations of}\;\; (p, \mu,\nu) \;\; (q, \alpha,\beta) \;\; (r, \gamma,\delta) 
%\nonumber
\end{eqnarray}
\end{subequations}
If $g_1 =  g_2$, we cannot recover the TT propagator from Eq. \eqref{eq33b} even if $\alpha\rightarrow 0$ \cite{Brandt:2007td}.

For the Bosonic ghost $\zeta^\mu$ we have a propagator and vertices that follow from Eqs. \eqref{eq27} and \eqref{eq28}.

%Eqs. (\ref{eq26}, \ref{eq27}) provide a complete set of Feynman rules for the 1EH action when the TT 
%gauge is used. It would be of interest to analyse the structure of divergences that aries in perturbation theory
%when using these Feynman rules.

%We now can consider some explicit calculations using these Feynman rules.

%\appendix*

%\section*{Appendix}

The arguments used in ref. \cite{faddeev:book1980,weinberg:book1995} can be used to show that when using a 
non-quadratic gauge fixing Lagrangian, physical results are independent of the gauge choice.

Beginning with the insertion of Eq. \eqref{eq7} into Eq. \eqref{eq6}, we have
\begin{eqnarray}\label{eqa1}
Z[\vj] &=& 
\int {\cal D} \vphi  \int {\cal D} \vth_1 {\cal D} \vth_2
\exp i\int \dd \left[ {\cal L}(\vphi) + \vj\cdot\vphi\right]
\nonumber \\
&&\delta(\unf\!{}_1(\vphi+\una\vth_1) - \vp_1)
\delta(\unf\!{}_2(\vphi+\una\vth_2) - \vp_2)
\nonumber \\
&&\det(\unf\!{}_1\una) \det(\unf\!{}_2\una) .
\end{eqnarray}
We can now insert into this equation a further factor of ``1'' 
\be \label{eqa2}
1 =  \int {\cal D} \vec\omega 
\delta(\unf\!{}_3(\vphi+\una {\bf \omega}) - \vec q)
\det(\unf\!{}_3\una) 
\ee
and then by interchanging ${\bf \omega}$ and $\vth_1$, and $\vp_1$ and ${\rm q}$
we see that 
$\unf\!{}_1$ and $\unf\!{}_3$ are interchanged without altering $Z[\vj]$, demonstrating that $Z$
is independent of the gauge fixing condition.

It would be interesting to derive a set of WTST and BRST identities associated with the gauge transformation of Eq. \eqref{eq4}
and the gauge choices of Eq. \eqref{eq13}.

\begin{acknowledgments}
We would like to thank CNPq  (Brazil) for a grant and Roger Macleod for encouragement.
\end{acknowledgments}

%\newpage

%\bibliographystyle{prsty}                                                                                          
%\bibliographystyle{unsrt} 
%\bibliography{all_new}       

\begin{thebibliography}{10}

\bibitem{Okubo:1979gt}
S. Okubo and Y. Tosa, Phys. Rev. {\bf D20},  462  (1979), [Erratum: Phys.
  Rev.D23,1468(1981)].

\bibitem{Buchbinder:1983ys}
I.~l. Buchbinder and I.~l. Shapiro, Yad. Fiz. {\bf 37},  248  (1983);
%
%\bibitem{Buchbinder:1985jc}
%I.~L. Buchbinder and I.~L. Shapiro, 
Acta Phys. Polon. {\bf B16},  103  (1985).

\bibitem{McKeon:1994ds}
D.~G.~C. McKeon, Can. J. Phys. {\bf 72},  601  (1994).

\bibitem{Kalmykov:1994fm}
M.~{\relax Yu}. Kalmykov, P.~I. Pronin, and K.~V. Stepanyantz, Class. Quant.
  Grav. {\bf 11},  2645  (1994).

\bibitem{Brandt:2015nxa}
F.~T. Brandt and D.~G.~C. McKeon, Phys. Rev. {\bf D91},  105006  (2015);
%
%\bibitem{Brandt:2016eaj}
%F.~T. Brandt and D.~G.~C. McKeon, 
Phys. Rev. {\bf D93},  105037  (2016).

\bibitem{Brandt:2007td}
F.~T. Brandt, J. Frenkel, and D.~G.~C. McKeon, Phys. Rev. {\bf D76},  105029
  (2007).

\bibitem{Brandt:2009qi}
F.~T. Brandt and D.~G.~C. McKeon, Phys. Rev. {\bf D79},  087702  (2009).

\bibitem{Brandt:2009rq}
F.~T. Brandt, J. Frenkel, D.~G.~C. McKeon, and J.~B. Siqueira, Phys. Rev. {\bf
  D80},  025024  (2009).

\bibitem{Brandt:2011zb}
F.~T. Brandt and D. McKeon, Phys. Rev. {\bf D84},  087705  (2011).

\bibitem{McKeon:2014iea}
D.~G.~C. McKeon, Can. J. Phys. {\bf 93},  1164  (2015).

\bibitem{McKeon:2010nf}
D.~G.~C. McKeon, Int. J. Mod. Phys. {\bf A25},  3453  (2010).

\bibitem{faddeev:book1980}
L.~D. Faddeev and A.~A. Slavnov, {\em Gauge fields. Introduction to Quantum
  Theory} (Benjamin Cummings, Reading MA, 1980).

\bibitem{weinberg:book1995}
S. Weinberg, {\em Quantum Theory of Fields II} (Benjamin Cummings, Cambridge,
  1995).

\bibitem{Feynman:1963ax}
R.~P. Feynman, Acta Phys. Polon. {\bf 24},  697  (1963).

\bibitem{DeWitt:1967yk}
B.~S. DeWitt, Phys. Rev. {\bf 160},  1113  (1967);
%
%\bibitem{Dewitt:1967ub}
%B.~S. DeWitt, Phys. Rev. 
{\bf 162},  1195  (1967);
%
%\bibitem{DeWitt:1967uc}
%B.~S. DeWitt, Phys. Rev. 
{\bf 162},  1239  (1967).

\bibitem{Faddeev:1967fc}
L.~D. Faddeev and V.~N. Popov, Phys. Lett. {\bf B25},  29  (1967).

\bibitem{Mandelstam:1968hz}
S. Mandelstam, Phys. Rev. {\bf 175},  1580  (1968).

\bibitem{Nielsen:1978mp}
N.~K. Nielsen, Nucl. Phys. {\bf B140},  499  (1978).

\bibitem{Kallosh:1978de}
R.~E. Kallosh, Nucl. Phys. {\bf B141},  141  (1978).

\end{thebibliography}

\end{document}